\documentclass[sn-mathphys]{sn-jnl}

\jyear{2021}%
\usepackage{pbox}

\theoremstyle{thmstyleone}%
\theoremstyle{thmstyletwo}%

\theoremstyle{thmstylethree}%

\begin{document}

\title[Article Title]{Human Digital Twin: A Survey}


\author[1]{\fnm{Yujia} \sur{Lin}}\email{M202120818@xs.ustb.edu.cn}

\author[2]{\fnm{Liming} \sur{Chen}}\email{l.chen@ulster.ac.uk}
\equalcont{These authors contributed equally to this work.}

\author[2]{\fnm{Aftab} \sur{Ali}}\email{a.ali@ulster.ac.uk}

\author[2]{\fnm{Christopher} \sur{Nugent}}\email{cd.nugent@ulster.ac.uk}

\author[2]{\fnm{Cleland} \sur{Ian}}\email{i.cleland@ulster.ac.uk}

\author[1]{\fnm{Rongyang} \sur{Li}}\email{ustblirongyang@sina.com}

\author[1]{\fnm{Dazhi} \sur{Gao}}\email{gaodz9803@163.com}

\author[1]{\fnm{Hang} \sur{Wang}}\email{wanghang146@126.com}

\author[1]{\fnm{Yajie} \sur{Wang}}\email{wangyajie9@163.com}

\author*[1]{\fnm{Huansheng} \sur{Ning}}\email{ninghuansheng@ustb.edu.cn}

\affil*[1]{\orgdiv{School of Computer and Communication Engineering}, \orgname{University of Science and Technology Beijing}, \orgaddress{\street{Xueyuan Street}, \city{Beijing}, \postcode{100083}, \state{Beijing}, \country{China}}}

\affil[2]{\orgdiv{School of Computing}, \orgname{Ulster University}, \orgaddress{\street{2-24 York Street}, \city{Belfast}, \postcode{BT15 1AP}, \state{Belfast}, \country{United Kingdom}}}



\abstract{Digital twin has recently attracted growing attention, leading to intensive research and applications. Along with this, a new research area, dubbed as "human digital twin" (HDT), has emerged. Similar to the conception of digital twin, HDT is referred to as the replica of a physical-world human in the digital world. Nevertheless, HDT is much more complicated and delicate compared to digital twins of any physical systems and processes, due to humans' dynamic and evolutionary nature, including physical, behavioral, social, physiological, psychological, cognitive, and biological dimensions. Studies on HDT are limited, and the research is still in its infancy. In this paper, we first examine the inception, development, and application of the digital twin concept, providing a context within which we formally define and characterize HDT based on the similarities and differences between digital twin and HDT. Then we conduct an extensive literature review on HDT research, analyzing underpinning technologies and establishing typical frameworks in which the core HDT functions or components are organized. Built upon the findings from the above work, we propose a generic architecture for the HDT system and describe the core function blocks and corresponding technologies. Following this, we present the state of the art of HDT technologies and applications in the healthcare, industry, and daily life domain. Finally, we discuss various issues related to the development of HDT and point out the trends and challenges of future HDT research and development.}

\keywords{Human Digital Twin, Generic Architecture, Human Modeling Technology, Digital Twin}



\maketitle

\section{Introduction}\label{sec1}

Under Industry 4.0, digital twin (DT) attracted increasing attention \cite{sharotry2020digital}. DT is the process of digital definition and modeling of the composition, characteristics, functions and performance of physical entities using information technology. It refers to information models that exist in the virtual digital world and are completely equivalent to physical entities, and is used to simulate, analyze and optimize physical entities \cite{tao2019make}. DT is wildly used in industry \cite{pires2019digital}\cite{opoku2021digital}, healthcare \cite{croatti2020integration}\cite{liu2019novel}, intelligent transportation \cite{rudskoy2021digital}\cite{sahal2021blockchain}, and other domains. With the rapid development of DT related technology (e.g., Internet of Things (IoT), 5G \textbackslash 6G, and artificial intelligence (AI)), researchers try to reconstruct the digital twin world in virtual cyberspace, extending from the application of atoms and devices to cells, hearts and human bodies. Along with this, a novel research area, dubbed as "human digital twin" (HDT), has emerged. Similar to the conception of DT, HDT is referred to as the replica of a physical-world human in the digital world.

HDT has shown considerable potential and attracted extensive attention from industry and academia since it was put forward. However, the development of HDT is still in its infancy. Despite the growing number of publications discussing HDT, it still can be found that the existing research lacks an in-depth analysis of HDT from the perspective of generic frameworks, technologies, applications, and challenges. Thus, we intend to analyze the status of HDT research, propose a comprehensive architecture to organize the HDTS, present the state of the art of core technologies and applications of HDT, and discuss various potential trends and challenges related to the development of HDT. The main contributions of this paper can be summarized as follows:

\begin{enumerate}[1.]
\item Present a comprehensive literature review to discuss the state-of-the-art of HDT to analyze underpinning technologies and establish typical frameworks for the core HDT functions or components.

\item Propose a comprehensive framework of HDT, and describe the core technology of using multi-modal and multi-source data to model human organ \textbackslash body and behavior to construct HDT.

\item Provide an expectation on the future trends and challenges of HDT, to promote further research.
\end{enumerate}

The overall structure of the paper takes the form of seven sections, including this Introductory Section. Section \ref{sec2} presents the related review works, including the concept and evolution of DT, and the emergence of HDT. Section \ref{sec3} investigates the state of the art of HDT. Section \ref{sec4} proposes a generic human digital twin system (HDTS) architecture. Section \ref{sec5} reviews the core technologies need to apply HDTS. The applications of HDT in healthcare, industry, and daily life are presented in Section \ref{sec6}. Section \ref{sec7} discusses various HDT trends and challenges from the perspectives of technology, social, and human thinking and cognition issues, and Section \ref{sec8} draws conclusions. For convenience, a summary of all relevant abbreviations is shown in Table~\ref{tab1}.

\begin{table}[ht]
\begin{center}

\begin{minipage}{290pt}
\tabcolsep=0.5cm 
\caption{Summary of abbreviations}\label{tab1}%
\begin{tabular*}{\textwidth}{@{\extracolsep{\fill}}ll@{\extracolsep{\fill}}}\toprule
\makebox[0.2\textwidth][l]{Acronyms} & \makebox[0.8\textwidth][l]{Description} \\ 
\midrule

DT  & digital twin \\
NASA  &  Aeronautics and Space Administration\\
PTC  & Parametric Technology Corporation\\
PLM  &  Product Lifecycle Management\\
MLM  &  Mirrored Space Model\\
IMM  & Information Mirroring Model\\
GE & General Electric Company\\
BHU  &  BeiHang University\\
HDT  &  human digital twin\\
HDTS  &  human digital twin system\\
PPG  &  photo plethyamo graph\\
AI &  artificial intelligence\\
DL &  deep learning\\
ML &  machine learning \\
LSTM&  long-short-term memory \\
CNN&  convolutional neural networks \\
SVD&  singular value decomposition \\
DNN&  deep neural networks \\
KNN&  k-nearest neighbor \\
ECOC&  error-correcting output code \\
NB&  Naïve Bayes \\
RF&  Random Forest \\
MLP&  Multilayer Perceptron \\
GNN&  graph neural network \\
MRI&  Magnetic resonance imaging \\
GAN&  generative adversarial network \\
IoT&  Internet of Things \\
TTS-CPS& Trusted Twins for Securing Cyber-Physical Systems  \\
DTN &  digital twin network \\

\botrule
\end{tabular*}
\end{minipage}
\end{center}
\end{table}

\section{Related Review Works}\label{sec2}
\subsection{The Concept and Evolution of Digital Twin}\label{subsec2.1}
Apollo program launched by the National Aeronautics and Space Administration (NASA) in the 1970s can be regarded as the germination of digital twin (DT), which is the first program to use the concept of “twins” \cite{piascik2010technology}. They constructed two identical space vehicles, the one on earth was used to predict the situation of the other one in space. Although this program is not a complete digital twin system, researchers realized the importance of constructing physical twins \cite{boschert2016digital}

\begin{figure}[ht]%
\centering
\includegraphics[width=0.9\textwidth]{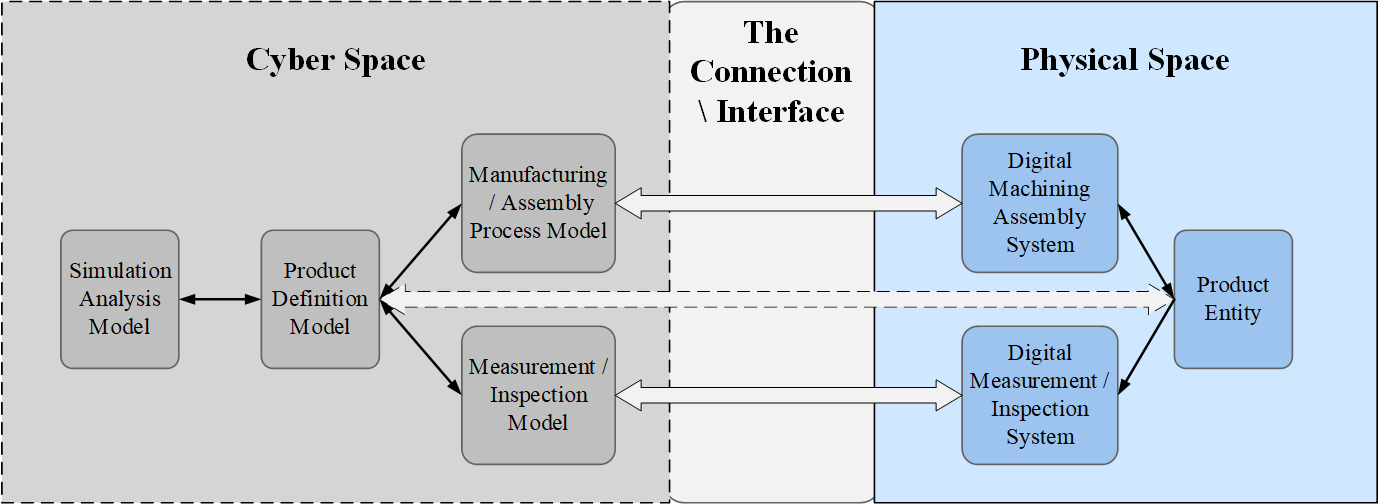}
\caption{Digital twin conceptual model}\label{fig1}
\end{figure}

In 2003, Grieves \cite{grieves2017digital} proposed the DT concept of “Virtual Digital Expression Equivalent to Physical Product” in his Product Lifecycle Management (PLM). However, this concept was not explicitly named DT at that time. Instead, it was referred to as the “Mirrored Space Model” (MLM) \cite{grieves2005product} from 2003 to 2005, and the “Information Mirroring Model” (IMM) \cite{githens2007product} from 2006 to 2010. Its conceptual model contained all core blocks of DT, including the physical space, the cyberspace, and the connection between the two spaces \cite{zhuang2017connotation}, as shown in  Figure \ref{fig1}. Compared with the “twin” put forward by the NASA Apollo program, Grieves has completed the transformation from entity to a digital model. Therefore, it could be seen as the origin of DT.

The development of various advanced technologies (e.g., Internet of things (IoT), cloud computing, edge computing, and artificial intelligence (AI)) provides technical support for the rapid application of DT. Beyond the DT conceptual model adopted for the aerospace domain, DT has also been widely applied to urban governance, transportation, manufacturing, energy, and other industries. Under the trend of Industry 4.0, especially in intelligent manufacturing, DT plays an important role in realizing the interactive integration of the virtual digital world and the physical world. Many well-known enterprises and organizations attach great importance to DT and have begun to explore a new mode of intelligent production based on DT. Table~\ref{tab2} provides the definitions of DT proposed by different enterprises or organizations, as well as their products.

\begin{table}[!ht]
\begin{center}
\begin{minipage}{\textwidth}
\caption{Definitions of Digital Twin in enterprises \textbackslash organizations}\label{tab2}%
\begin{tabular}{@{}lll@{}}
\toprule
 \begin{tabular}[c]{@{}l@{}} Enterprises \textbackslash \\ Organizations \end{tabular}  & \multicolumn{1}{c}{Definitions} & Products \\
\midrule

 \begin{tabular}[c]{@{}l@{}} National \\ Aeronautics \\ and Space \\ Administration \\ (NASA) \end{tabular}  
 &	 \begin{tabular}[c]{@{}l@{}}  \pbox{7.5cm} {“An integrated multi-physics, multi-scale, probabilistic simulation of a vehicle or system that uses the best available physical models, sensor updates, fleet history, etc.” \cite{piascik2010technology} } \end{tabular}
 & \begin{tabular}[c]{@{}l@{}} Flying \\ Twin \end{tabular} \\


  
Siemens	& 
\begin{tabular}[c]{@{}l@{}} \pbox{7.5cm} {The concept of "integrated digital twin" was proposed by Siemens, which included the accurate continuous mapping and progressive relationship between digital twin products, digital twin production and digital twin operation, so as to achieve the ideal delivery of high-quality products finally. Based on the consistent data model in all aspects of the product life cycle, some practical operations can be simulated accurately and indeed. \cite{Siemens} }\end{tabular}
& \begin{tabular}[c]{@{}l@{}} Siemens \\ PLM \end{tabular} \\

  \specialrule{0em}{3pt}{3pt}
  

  \specialrule{0em}{3pt}{3pt}

 \begin{tabular}[c]{@{}l@{}} General \\ Electric \\ Company \\ (GE) \end{tabular}	
& 
\begin{tabular}[c]{@{}l@{}} \pbox{7.5cm} {“Digital Twin is most commonly defined as a software representation of a physical asset, system or process designed to detect, prevent, predict and optimize through real-time analytics to deliver business value.” \cite{GE} }\end{tabular}
& \begin{tabular}[c]{@{}l@{}} Predix \end{tabular} \\

  \specialrule{0em}{3pt}{3pt}
  
Dassault & 
\begin{tabular}[c]{@{}l@{}} \pbox{7.5cm} {“A Digital Twin is an executable virtual model of a physical thing or system. The physical thing can be anything from a manufactured object, every single product has definable characteristics in the real world, and the Digital Twin combines and portrays the attributes virtually.”\cite{graupner20203dexperience}}\end{tabular}
& \begin{tabular}[c]{@{}l@{}} 3DEXPE \\ -RIENCE \\ Twin \end{tabular} \\

  \specialrule{0em}{3pt}{3pt}

 \begin{tabular}[c]{@{}l@{}} BeiHang \\ University \\ (BHU)\end{tabular}	
& 
\begin{tabular}[c]{@{}l@{}} \pbox{7.5cm} {Tao et al. \cite{tao2019make} put forward a five-dimension digital twin model comprising a physical entity, virtual entity, services, digital twin data, and connection. Based on the five-dimension model, typical applications in different fields were explored.}\end{tabular}
& \begin{tabular}[c]{@{}l@{}} Five\\ -Dimension \\ Digital \\ Twin \\ Model \end{tabular} \\

\botrule
\end{tabular}
\end{minipage}
\end{center}
\end{table}

Table~\ref{tab2} lead us to the conclusion that most previous work and applications focused on the discussion of the product lifecycle. Only some of them consider the whole human lifecycle \cite{shengli2021human}. With the continuous development of DT, research began to emphasize putting human first and taking the entire lifecycle of human beings into consideration, which received more and more attention \cite{talkhestani2019architecture}. People considered the entire lifecycle of human beings when designing and using DT models. DT models designed for this purpose were highly specialized, and provided an excellent human-machine interface to help people in the physical world work better. For example, Wang et al. \cite{wang2020digital} designed an intelligent interaction welding and welder behavior analysis platform, collected and digitized the real welding work, and provided personnel training in the form of Virtual Reality (VR). A set of offshore DT platforms were created by Pairet et al. \cite{pairet2019digital} for training and testing in the offshore energy industry, assisting people in learning how to use remotely operated tools for automated energy extraction. Martinez-Velázquez et al. \cite{martinez2019cardio} simulated the human heart for ischemic heart disease. Subsequently, the term DT has been extended to humans using the term “human digital twin (HDT)”.

\subsection{The Emergence and Vision of Human Digital Twin}\label{subsec2.2}
Human digital twin (HDT) is an incarnation in the digital world of a human in the physical world. Shengli et al. \cite{shengli2021human} defined it as a model or database which records human current and historic data. Firstly, human information obtained continuously through various smart sensors (e.g., wearable devices, smartphones, and GPS), is transferred to the digital world consecutively. Secondly, the database is updated according to the recorded information. Thirdly, the HDT model analyzes the current and historic data to extract meaningful insights. Finally, it provides feedback information (e.g., diagnoses, predictions, and other suggestions) to the human. Miller et al. \cite{miller2022unified} further refined the definition of HDT. They pointed out that a human in the physical world may be an individual, who could be characterized as a model, or a group of humans with common attributes, which could be summarized into eight categories of human attributes, including physical, physiological, perceptual performance, cognitive performance, personality characteristics, emotional state, ethical stance, and behavior. The HDT they defined includes both first principles models which are based on fundamental understanding and statistical models, which are both modeled using various attributes from an individual or a human class \cite{lewis2019reference}. Recently, HDT are widely used in various domains, including precision medicine \cite{liu2019novel}\cite{chakshu2021towards}\cite{shamanna2021retrospective}\cite{ahmed2021potential}\cite{mourtzis2021smart}, fitness management \cite{barricelli2020human}, industrial production \cite{locklin2021architecture}\cite{greco2020digital}\cite{sun2020digital}\cite{al2020user}\cite{sharotry2020digital}\cite{ariansyah2020towards}\cite{montini2021meta}\cite{baskaran2019digital}.

From the above review, it can be concluded that HDT is a branch of DT and derives from it. However, there are various differences between HDT and DT, which mainly present from the aspects of data and HDT model design.

Regarding data, it is more challenging for researchers to select, collect, and protect the various data for the HDT model. First of all, human avatars are more difficult to describe accurately than machines. When choosing the data to construct a human avatar, researchers need to consider not only the inner characteristic (e.g., physiological signal, mood and emotion) and external characteristics (e.g., height, weight and gender) of human, but also the human social attributes (e.g., ethics and privacy, interpersonal network and human social interaction), as well as the influence of the environment on human. Moreover, human also has the characteristics of heredity and variation. It can be seen that the construction of HDT requires various unique data structures to merge these attributes, which makes it challenging to construct HDT \cite{shengli2021human}. Secondly, unlike machines, a human does not include integrated sensors, which causes the data used in HDT modeling and optimization cannot be continuously collected and transmitted to the digital world \cite{lauer2022designing}, which may lead to data gaps and cause wrong decisions and affect the accuracy of HDT prediction results \cite{barricelli2020human}. Finally, HDT modeling uses a large amount of human-related data, and a considerable proportion of data is directly related to human privacy \cite{de2020digital}. Therefore, paying more attention to the privacy of HDT is necessary. People’s trust in HDT is crucial to HDT applicability \cite{stacchio2022will}, especially in the medical domain \cite{petrova2020digital}.

Regarding the HDT model design, the HDT model is more complicitly designed than DT due to the consideration of the entire human lifecycle and the distinctive characteristics of humans in the design of HDT. Firstly, the behaviors allowed by HDT model must be based on human prototypes in the physical world, so it can not only perform professional behaviors as DT in work, but also do what humans can do in daily life, and have continuous learning capabilities according to changes in the environment and themselves \cite{locklin2021architecture}. Secondly, HDT designers can assign one or more behaviors to the HDT model as required, and need to provide extensible interfaces for unassigned behaviors \cite{shengli2021human}. Thirdly, the reusability of the DT model differs from that of the HDT model. The HDT model can adapt by learning current data from the physical world and the historical data stored in the digital world, making it more suitable for the current scene. It is like a process of continuous human learning and growth, through the learning of different behaviors, from "general" to "professional" \cite{locklin2021architecture}.

\section{The State-of-Art of Human Digital Twin}\label{sec3}
\subsection{Overview of the Literature}\label{subsec3.1}

With the increasing importance of digitalization and the trends of IoT, big data, cloud computing, edge computing and AI, more researchers began focusing on HDT. The popularity of HDT continues to increase, and the application of HDT has great potential in engineering, aviation, education, medicine and a wide variety of other domains. 

We have searched HDT-related publications based on the search field, as shown in Table \ref{tab3}. “Human digital twin” is the primary term used to search. HDT has other names in different domains of application. For instance, it is sometimes referred to as the “digital twin of patients” in healthcare, the “digital twin for employees” in industrial production or the “digital athlete” in sports health. To make a comprehensive review of HDT, we include these terms in the search terms.


\begin{table}[ht]
\begin{center}

\begin{minipage}{\textwidth}
\tabcolsep=0.5cm 
\caption{Search field of the literature review}\label{tab3}%
\begin{tabular*}{\textwidth}{@{\extracolsep{\fill}}ll@{\extracolsep{\fill}}}\toprule
\makebox[0.1\textwidth][l]{Search Filed} & \makebox[0.8\textwidth][l]{Content} \\ 
\midrule

Database &	\begin{tabular}[c]{@{}l@{}}
Scopus, Google Scholar, Web of Science, IEEE Xplore, Springer, and \\ Elsevier, etc.
\end{tabular}\\
  \specialrule{0em}{1pt}{2pt}
  
Search Term & \begin{tabular}[c]{@{}l@{}}
“Human Digital Twin”, “patient Digital Twin”, “Digital Twin for \\ mental”, "Digital Twin in human", "digital athlete", " digital twin for \\ employees", " digital twin as human representation", etc.
\end{tabular}\\
  
  \specialrule{0em}{1pt}{2pt}
  
Search Scope & Title, Abstract, Key World \\
  
  \specialrule{0em}{1pt}{2pt}
  
Time Scope	& 2018.02 -2022.06 \\

\botrule
\end{tabular*}
\end{minipage}
\end{center}
\end{table}

\begin{figure}[ht]%
\centering
\includegraphics[width=0.6\textwidth]{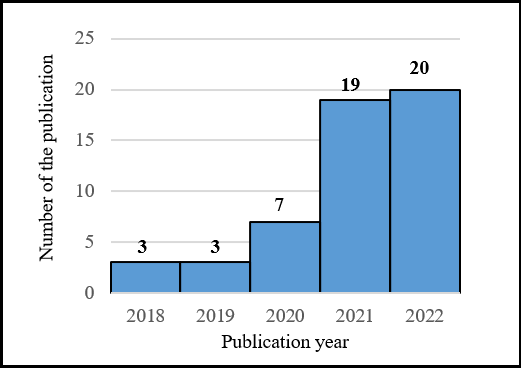}
\caption{Publication year for paper in the literature review}\label{fig2_1}
\end{figure}

In order to establish a clear structure of existing literature we decide to clarify publications based on the nature of the papers, namely the concept, technology, framework or application of HDT. After excluding irrelevant publications, 56 literatures with high relevance were screened. Figue~\ref{fig2_1} shows a steady increase in the number of HDT publications. The growth of the number of HTD-related publications began recently, after 2018. Before 2020, the development of HDT in the academic field is relatively slow. From 2020 to 2022, the number of HDT publications in the academic field increased rapidly. It shows that HDT is gradually coming out of the embryonic stage and entering the rapid development stage. Researchers have shifted from exploring the concept of HDT to exploring the generic framework of HDT and the practical application and related technologies.

HDT originated from adding human resources in the industry domain. The earliest publications did not mention the term “human digital twin”, instead using other terms like “digital twin for employees” or “digital athlete”. Various human attributes were used to develop DT models for employees, which is helpful to the coordination between workers and the system. Graessler et al. \cite{graessler2017integration} developed the DT, combining the integration of human resources with the task assumption of computer systems, communicating and coordinating between employees and production systems, and acting as the representative of employees in the digital world.  the authors optimized the DT for employees \cite{graessler2018intelligent}, used it for the task of assembly station, and demonstrated the impact of human integration in the production system and production control system. HDT not only appeared in the industry domain, but also emerged from the sports domain. Garon et al. \cite{garon2018reordering} overviewed the virtual world and the legal framework regulating content ownership, and discussed the legal issues of avatar actors and digital athletes in the sports domain.  Baskaran et al. recognized assembly operations from real-world vehicle assembly factories and created digital twin models of the human body in the Siemens Tecnomatix suite, and this is where the term "human digital twin" first appeared in the literature \cite{baskaran2019digital}. This research started with simulating various human body models to find the limitations of performing assembly tasks based on gender, weight and height. Furthermore, the research also introduced a mobile robot DT, which provided help for human in the physical world to perform assembly operations. Finally, the research evaluated the results from process time and joint ergonomics, and revealed the limitations of the combined DT modeling of human and robot cooperation.

\subsection{Categorization of the Literature}\label{subsec3.2}
The organization of HDT is categorized as follows: 

\begin{itemize}
\item Concept, technology and conceptual model category
\item Framework with application model category
\item Discussion of open issues category
\end{itemize}

Table~\ref{tab4}. shows the detailed description of each category and typical examples of each literature classification.

\begin{table}[!ht]
\begin{center}
\begin{minipage}{\textwidth}
\caption{Organization Way of the literature review}\label{tab4}%
\begin{tabular}{@{}lll@{}}
\toprule
 \begin{tabular}[c]{@{}l@{}} Classification \\ of the existing \\ organization \\ way \end{tabular}  & \multicolumn{1}{c}{Detailed description} & \multicolumn{1}{c}{Typical example} \\
\midrule

  \specialrule{0em}{3pt}{3pt}
  
 \begin{tabular}[c]{@{}l@{}} Concept, \\ technology \\ and \\ conceptual \\ model\end{tabular}  
 &	 \begin{tabular}[c]{@{}l@{}}  \pbox{3cm} {Literature in this category focuses on concepts, definitions, and capabilities of HDT, or proposes a general conceptual model of HDT applications in respective domains, with no in-depth analysis and no detailed case study.} \end{tabular}
 &  \begin{tabular}[c]{@{}l@{}}  \pbox{6cm} {(1)	The authors in \cite{shengli2021human} presented HDT following an explanation of the history and uses of DT, provided the concept, conceptual model, and characteristics of HDT, and contrasted HDT with DT. Finally, an HDT was built using the characteristics of the information flow chart, the architecture, and the implementation strategy of HDTS. \\
(2)	The authors of \cite{miller2022unified} organized a review of the literature and investigated the modeling structure and use cases. They then provided a structural definition of an HDT through a block definition diagram, defined HDT and HDTS, and discussed use cases for HDTS.} \end{tabular} \\

   \specialrule{0em}{3pt}{3pt}
   
 \begin{tabular}[c]{@{}l@{}} Framework \\ with \\ application \\ model \end{tabular}	& 
\begin{tabular}[c]{@{}l@{}} \pbox{3cm} {Literature in this category describes HDT’s enabling technology in detail and gives general descriptions of HDT applications in respective domains.}\end{tabular}
 &  \begin{tabular}[c]{@{}l@{}}  \pbox{6cm} {(1)	The authors of \cite{wang2022human} proposed a framework of HDT-driven human-cyber-physical systems (HCPS). The framework included human, the physical system, and the cyber system, and the human is the center of HDTS. Various enabling technologies (e.g., sensing technology, computing and analysis technology, and control and interaction technology) were investigated to achieve this framework.\\
(2)	The authors of \cite{ferdousi2021iot} proposed a Digital Twin of Mental Stress (DTMS) model that employed IoT-based multi-modal sensing and ML for mental stress prediction, achieved 98\% accuracy, and concluded that the optimal Digital Twin Features (DTF) could shorten the classification time.} \end{tabular} \\

  \specialrule{0em}{3pt}{3pt}
  
 \begin{tabular}[c]{@{}l@{}} Discussion \\ of open \\ issues \end{tabular}	
& 
\begin{tabular}[c]{@{}l@{}} \pbox{3cm} {Literature in this category mainly focuses on the open issues of HDT, such as whether a HDT could represent the human in ethics \cite{braun2021represent}, or the law of digital athletes \cite{lauer2022designing}.}\end{tabular}
 &  \begin{tabular}[c]{@{}l@{}}  \pbox{6cm} {The authors of \cite{lauer2022designing} introduced the methodology to overview the relevant recent work, distilled vital design considerations of HDTS in regulation and ethical considerations, transparency and trust, dynamism and flexibility and behavior and cognitive considerations.
} \end{tabular} \\

  \specialrule{0em}{3pt}{3pt}

\botrule
\end{tabular}
\end{minipage}
\end{center}
\end{table}

\subsection{Typical Frameworks of Existing HDT}\label{subsec3.3}
Through the above research, we discuss the organization of three related publication categories, including the framework with the application model category. Next, the structure of HDT and their application in system design, development and operation will be presented in this subsection. The existing frameworks of HDTS could be analyzed from the following two typical examples.

Firstly, Michael et al. \cite{miller2022unified} presented the HDTS framework mainly through block definition diagrams. The first diagram showed the primary components of a DTS, dividing it from DT, real-world twin and interchange components. The second diagram described an exploded view of a generic HDT, giving the first diagram's sub-block. The other diagrams explained the HDTS use cases from product development and acquisition-based and operations and sustainment use cases based on literature. It is a universal HDTS framework, which conforms to the conceptual model proposed by Greeves \cite{zhuang2017connotation}. The contributions of this HDTS framework are; first, it proposed that human in the physical world can be an individual or a group of humans with common characteristics, and HDT can not only exist as a mathematical representation of an individual or a group of humans, but also exist as a virtual entity and appear in the virtual or real-world system. Secondly, it regarded HDT as the first principal or statistical model based on fundamental understanding and introduced the relevant attributes of humans for modeling in detail. In general, this framework is a general framework that can be used as the basis to apply to various domains.

Secondly, the HDTS framework proposed by Baicun Wang et al. \cite{wang2022human}, named HDT-driven HCPS framework, shows three pillars, including human, the physical system, and the cyber system. In addition, it has developed the original HDT into a broader concept, and added the physical representation and virtual model of human-cyber-physical (HCP) elements. Human is at the center of HDTS and coordinate HCP to drive HDT. Considering the requirements of industrial production, the physical representation of HDT is integrated into the physical system, that is, it is a combination of physical entities (i.e., intelligent sensors and actuators) and the environment. Physical systems help human interact with the cyber system by receiving remote operations. Cyber system receives not only human decision-making knowledge, but also supports human knowledge learning, representation and organization. It is not only used to monitor human and physical systems, but also to manage the virtual model of HDT, that is, the virtual model of human and the virtual model of the physical system. It supports the critical decision of continuously improving HDT-driven HCP using optimization and artificial intelligence. Secondly, it also shows that data (including decision-making and signals) flows between the physical system and cyber system, human and physical system, and human and cyber system in real-time. In short, it extends the scope of HDTS from human, HDT, and interactive components to human and physical systems and their physical representations in the physical world, and cyber system, virtual models of humans and virtual models of the physical system in the digital world. It reflects the interaction between the HDT model and human and their environment and has great applicability in industrial production.

\section{A Generic System Architecture of Human Digital Twin}\label{sec4}

Grieves et al. \cite{zhuang2017connotation} divided the conceptual model of DT into three core blocks: the physical world block, the digital world block, and the connection and interface block to connect the two worlds. HDT is a branch of DT, and they are essential to building an entirely consistent digital model in the digital world for physical objects in the physical world. Therefore, we attempt to define the human digital twin system (HDTS) by extending a way proposed in literature \cite{miller2022unified}.

\begin{figure}[ht]%
\centering
\includegraphics[width=0.9\textwidth]{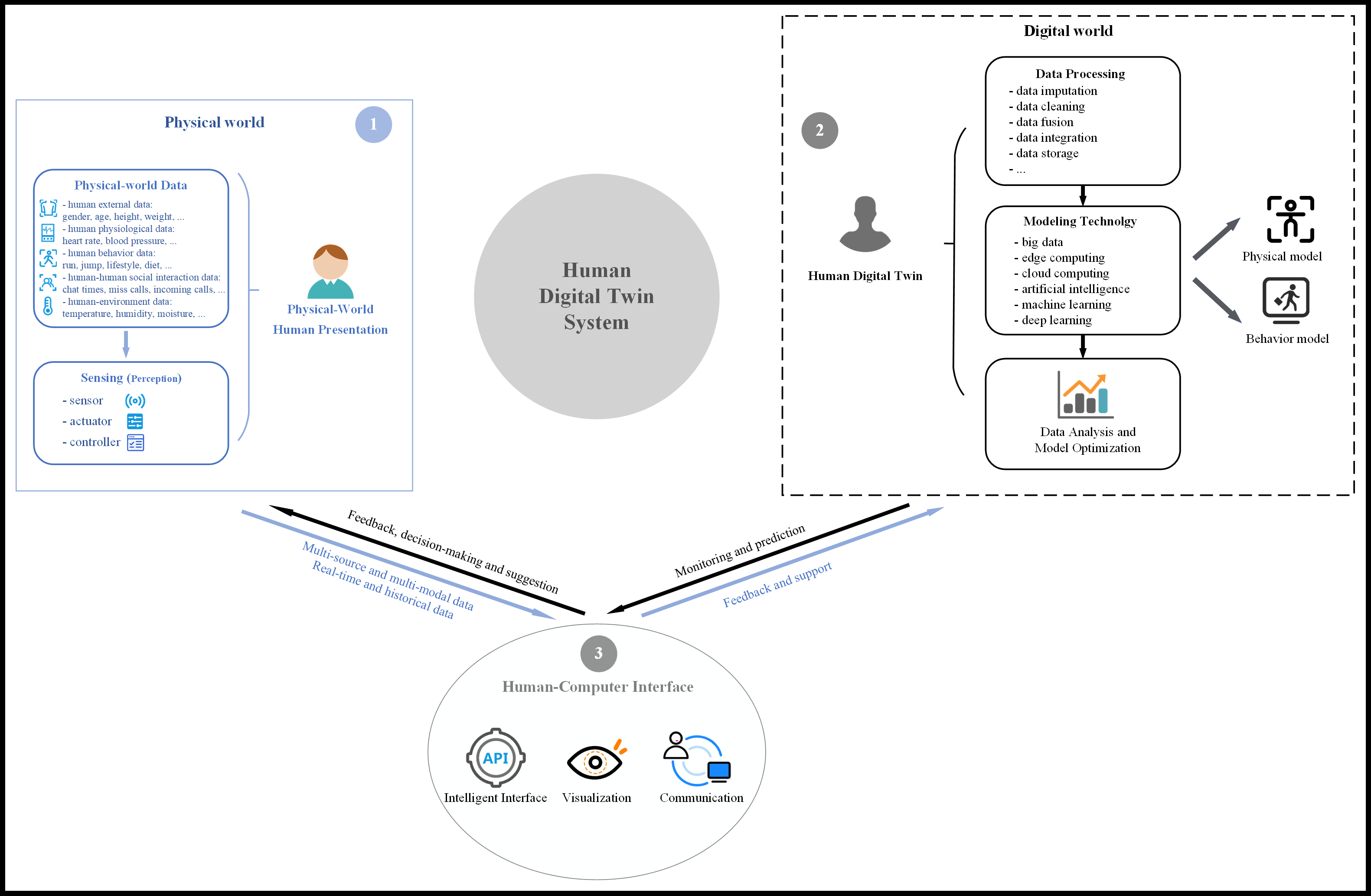}
\caption{The generic human digital twin system framework}\label{fig2}
\end{figure}

HDTS is a valuable tool to describe, monitor and simulate human in the physical world. Based on the block definition diagram proposed by literature [18], we design HTDS framework and refine the definition. Figure \ref{fig2} shows the HDTS designed by us, the division of the sub-blocks, as well as a correlation between each block.

The first block presents a physical-world human representation, including a data collection sub-block and a sensing and perception sub-block, which realizes the data collection function in the physical world. Choosing suitable human-related data for HDT modeling is critical. Human-related data can be divided into eight categories: human external data, physiological data, human-to-human social interaction data and human-to-environment data, the detailed descriptions and examples of these data are shown in Table~\ref{tab5}.

\begin{table}[!ht]
\begin{center}
\begin{minipage}{\textwidth}
\caption{The detailed descriptions and examples of HDT data}\label{tab5}%
\begin{tabular}{@{}ccl@{}}
\toprule
\multicolumn{1}{c}{Aspect}  & \multicolumn{1}{c}{Example} & \multicolumn{1}{c}{Description} \\
\midrule

  \specialrule{0em}{3pt}{3pt}
  
 \begin{tabular}[c]{@{}l@{}} \pbox{1.5cm}{ Human external data }\end{tabular}  
 &	 \begin{tabular}[c]{@{}l@{}}  \pbox{2cm} {Gender, age, height, weight, belly thickness, size, etc.} \end{tabular}
 &  \begin{tabular}[c]{@{}l@{}}  \pbox{7.5cm} {The body modeling of HDT needs to use the human external data of the natural human body \cite{sengan2022cost}. The data for building the 3D model can be obtained through 3D laser scanning technology \cite{kim2017motion}, visual recognition \cite{afzal2014rgb}, and depth cameras like Microsoft Kinect \cite{li2017dynamic}.} \end{tabular} \\

   \specialrule{0em}{3pt}{3pt}
   
 \begin{tabular}[c]{@{}l@{}} \pbox{1.5cm}{ Human \\ physiological \\ data} \end{tabular}  
 &	 \begin{tabular}[c]{@{}l@{}}  \pbox{2cm} {Heart rate, blood pressure, galvanic skin response (GSR), skin conductivity (SC), skin temperature (ST), etc.} \end{tabular}
 &  \begin{tabular}[c]{@{}l@{}}  \pbox{7.5cm} {When using 3D reconstruction technology to create human internal organ models, human physiological data is frequently used as a starting point. According to the laws of the relevant physiological data of reconstructed human organs, the HDT in the digital world is corrected by the physiological data from the physical world individual organs \cite{martinez2019cardio}\cite{mazumder2019synthetic}\cite{chakshu2021towards}\cite{crea2020focus}. Human physiological data can also be used to predict higher levels of behavior. For example, muscle tension can be used to predict worker fatigue \cite{sharotry2020digital}.} \end{tabular} \\
 
  \specialrule{0em}{3pt}{3pt}

\begin{tabular}[c]{@{}l@{}} \pbox{1.5cm}{Human behavior data}\end{tabular}  
 &	 \begin{tabular}[c]{@{}l@{}}  \pbox{2cm} {Run, jump, walk, lifestyle, diet, sleep hour, daily phone usage, etc.} \end{tabular}
 &  \begin{tabular}[c]{@{}l@{}}  \pbox{7.5cm} {Human behavior is usually used as behavior modeling together with data from other sources, which can be used to diagnose patients' physical and psychological diseases in healthcare domain \cite{ferdousi2021iot}\cite{chen20223d}, and to evaluate the cognitive performance of workers (e.g., knowledge, skills, ability, workload level, decision-making ability) in the industry domain \cite{sharotry2020digital}.} \end{tabular} \\
 
  \specialrule{0em}{3pt}{3pt}

\begin{tabular}[c]{@{}l@{}} \pbox{1.5cm}{Human-human social interaction data}\end{tabular}  
 &	 \begin{tabular}[c]{@{}l@{}}  \pbox{2cm} {Social media chat time, social media active time, missed calls, incoming calls, etc.} \end{tabular}
 &  \begin{tabular}[c]{@{}l@{}}  \pbox{7.5cm} {Due to human social characteristics, HDT modelling must take human social interaction data into account. Human social interaction in the physical world is expressed as HDT and HDT interaction in the digital world.} \end{tabular} \\
 
  \specialrule{0em}{3pt}{3pt}

\begin{tabular}[c]{@{}l@{}} \pbox{1.5cm}{Human-environment data}\end{tabular}  
 &	 \begin{tabular}[c]{@{}l@{}}  \pbox{2cm} {Temperature, humidity, cleanliness, illumination, etc.} \end{tabular}
 &  \begin{tabular}[c]{@{}l@{}}  \pbox{7.5cm} {The environmental issue also has an impact on HDTS. Because in the prediction process, HDT models in the virtual world need to make predictions according to the physical-world human representation environment. This issue is less apparent in literature, because the concept of DT is applied to the controlled environment \cite{miller2022unified}, and the HDTS does not necessarily need a broad environmental model. It only needs a model that reflects the information, energy and matter exchanged between human representation and the environment. Therefore, creating a standard environment model is unnecessary, making HDTs too complicated.} \end{tabular} \\
 
  \specialrule{0em}{3pt}{3pt}

\botrule
\end{tabular}
\end{minipage}
\end{center}
\end{table}

The use of these data helps to describe HDT more comprehensively. Each human and his corresponding HDT would change synchronously. All the human-related data changes (e.g., human external data, physiological data, behavior data, social interaction data, and environment data) in the physical world will be sensed and transmitted to the digital world, and the HDT will change accordingly. In order to realize cognitive sharing and real-time interaction between the physical and digital worlds, the sensor is responsible for collecting data from humans in the physical world and connecting to the environmental system where people live for environmental cognition (e.g., humidity and gas concentration). By fully capturing the attributes and states of the physical world to connect the physical world and the digital world, the sensing component sends data to the ingestion program of the digital world. Therefore, in sensing and perception, at least one sensor is needed, and the arithmetic unit and controller are optional components in HDTS, which may only exist in some HDTS.

The second block researches the construction of HDT in the digital world, and mainly contains data processing, modeling, simulation and analysis engine. First, in the data processing sub-block, data cleaning, data storage, data fusion and integration are necessary for data processing. As for data cleaning, HDTS connects the data recorded by the sensor through its physical world human representation. There are usually existing problems of data discontinuity and deletion. In this case, applying data interpolation technology to fill the missing data according to the structure of the basic dataset has proved to be an effective technology for processing biological data. Among them, the KNN interpolation algorithm \cite{hastie1999imputing}\cite{troyanskaya2001missing}\cite{speed2003statistical} is proved to be effective when applied to impute the missing values \cite{hastie1999imputing}\cite{troyanskaya2001missing} of the gene expression array. Research indicated that KNN interpolation could not only obtain higher performance, but also outperform other interpolation methods (e.g., case substitution, mean and mode, hot deck and cold deck, and interpolation methods using prediction models) \cite{batista2002study}. As for data storage, the physical world has a corresponding HDT stored in the digital world. Each of the HDT has a unique index, which can be used as its ID and can also be used as an account to log in to the HDT. Latift et al. \cite{latif2020case} regarded the data storage process as a part of HDTS in the design of HDTS. In addition, as for data fusion and integration, HDTS aims to fuse information from these sensors associated with the physical world human representation and integrate new real-time data with historical data \cite{tao2019make}. Secondly, modeling and simulation are the key sub-block of building HDT in the digital world. Human modeling and simulation create a virtual model by defining and extracting the key features of human to reflect the key characteristics and dynamic motion of human representation. The simulation further explores the system performance based on various modeling techniques \cite{miller2022unified}. The modeling methods we explored are divided into organ and body modeling at the physical level, human activity modeling, social interaction modeling and lifestyle modeling at the behavioral level. As the core sub-block of HDT, and even the critical functional module of HDTS, the modeling technology will be introduced in detail in Section IV. Finally, the data analysis and model optimization sub-block aims to further predict, evaluate and optimize the HDT through the analysis engine, makes a distortion for suggestions and decision-making through the prediction engine, and compares the prediction results with the data in the database, evaluates the performance of HDTS though the evaluation engine, determines whether the prediction results can be used as feedback to physical-world human representation. The optimization objective is determined according to the evaluation results to optimize the HDT further and improve its accuracy \cite{miller2022unified}.

The third block focuses on the human-computer interface. The intelligent interface is the bridge between the physical world and the digital world, and is a critical sub-block to realizing the real-time interaction between human representation and HDT. Therefore, the two sub-modules of intelligent interface and communication tools are necessary for the human-computer interface block. The visualization engine sub-block could strengthen the interaction between users and HDT, and enhance users’ understanding and trust of the HDT in the virtual world, which is realized through VR \textbackslash AR.

In order to clarify the relationship between the three blocks (physical-world human representation block, digital-world human digital twin block and human-computer interface block) and the information flow process in HDTS, we present a further explanation for this HDTS framework to make it more relevant, interactive and transparent. The working process of HDTS is as follows:

\begin{enumerate}[1.]
\item Determine the application purpose of HDT and select appropriate attributes. A different application of HDTS determines different data used in HDT modeling.

\item The sensors of the physical world perceive the relevant attributes and state of human and the surrounding environment in the physical world. The physical-world human representation is built through human-related data.

\item The intelligent interface of human-computer interaction connects the physical and digital worlds. It transfers the sensed multi-source and multi-modal data from the physical world to the digital world, providing support for HDT modeling in the digital world. Moreover, when the state of people (e.g., emotion, psychology, and behavior) in the physical world changes, the real-time data is also feedback to the digital world to optimize and modify the HDT model.

\item After receiving the data, the virtual world processes the data, selects features, and stores the processed current and historical data in the HDTS database.

\item HDTS uses modeling technology (e.g., ML, DL and continual learning) and the data in the database to build the HDT model corresponding to human in the physical world. HDT model can be divided into physical model and behavior model according to its state. Physical models are used to model organs or the human body. Among them, the organ model is mainly used for precision medical treatment, and the body model can be used for identity authentication, virtual shopping, and others. The behavior model is more widely used in industry, healthcare, daily life and other domains.

\item HDTS permits the HDT model to simulate and evolve with data collected in the physical world to provide the future prediction of corresponding individuals, conduct data analysis and evaluation, judge the accuracy of HDT model prediction, and adjust and optimize the HDT model.

\item The human-computer interaction interface module could provide a user-friendly HDTS interaction interface through visualization technology, and provide feedback.

\end{enumerate}

Compared to other latest HCPS frameworks (e.g., HDTS designed by \cite{miller2022unified}, and HCPS presented by \cite{wang2022human}), the HDTS, we proposed, surpasses the existing research in two aspects:

\begin{itemize}
\item With the development of various technologies (e.g., 5 \textbackslash 6G, edge computing, cloud computing and IoT), multi-modal and multi-source data can be applied to this framework, considering not only human data (external and physiological), but also the data generated by human social interaction, and also considers environment data which could impact HDT modeling. These data provide more features for the construction of HDT through mutual support, supplement and correction, and avoid the problem of single data affecting the prediction accuracy of HDT.

\item We consider the modeling of HDT from two aspects: physical (organs and body) and human behavior (human activities, social interaction and lifestyle). Compared with the framework proposed to apply in specific domains (e.g., industry, healthcare and sports), this framework increases the flexibility of HDT application and improves the universality of HDTS framework. 
\end{itemize}

\section{Core Human Digital Twin System Technology Review}\label{sec5}
One of the contributions of this paper is the proposed HDTS framework with the sensing (perception) technology and two modeling technologies of organs\ body and behavior. Therefore, we investigate the state of the art of these three core HDT-related technologies, and present each technology's straightforward methods, development status, advantages and disadvantages in the functional block of HDTS in this section.
\subsection{Sensing \textbackslash Perception}\label{subsec5.1}
Sensing is the basis of HDTS. The sensor is used to sensing the properties, parameters and actions of the entities and their surrounding in the physical world \cite{shengli2021human}, and efficiently provides data to the HDT of the design object \cite{erol2020digital}. The ubiquitous data acquisition technology has dramatically improved the perception ability of the physical world. The development of advancements in hardware (e.g., GPS, smartphone, wearable devices) has made the modeling and analysis of HDT possible \cite{wang2022human}.

The realization of HDT requires multi-sources data. The multi-sources include but are not limited to sensors, databases, and social media. Wearable sensors collect signals from the human and provide good feedback. It can effectively obtain human physiological data (e.g., galvanic skin response (GSR), skin conductivity (SC), skin temperature (ST), skin blood flow volume (BVP),). Sensors not only collect data of human living environment (e.g., temperature, movement, humanity, distance), but also can automatically collect pictures or videos of human facial expressions, body language and social behaviors. In addition, social media data collection obtains social text, pictures, user comments, likes and other information, from the website public application programming interface of social networking websites. Moreover, medical databases contain clinical monitoring data, early warning of chronic disease data, daily activity data, and vital sign detection data.

Multi-source and multi-modal data can provide more information than single data \cite{zheng2015methodologies}. Through mutual support, supplements and correction, it can provide more accurate information and meet the different HDT modeling requirements. It's predicted that the HDT model using advanced sensing technology will exactly predict the physiological and psychological status of human beings \cite{wang2022human}. Research extracts features from the data collected by the sensing technology for HDT modeling. These data provide the possible prediction of the physical world. Moreover, the sensing or perception technology used to collect data for HDT modeling is quite different for different domains.

One of the reasons HDT was introduced in healthcare is so that a thorough and ongoing examination of a person's health and personal history could be done in the context of healthcare. This would enable the diagnosis or treatment of illnesses, as well as the prediction of their occurrence \cite{bruynseels2018digital}. Liu et al. \cite{liu2019novel} proposed a conceptual framework, and described the method of using data collected by IoT-connected wearable devices to construct the HDT. HDTS performed well in predicting the performance of athletes by collecting current fitness related data (e.g., exercise time, sleep time and quality, diet) \cite{barricelli2020human}. Another example comes from Romero et al. \cite{romero2016towards}, who designed a human operator 4.0 in the context of the industrial domain using physical, sensory, and cognitive performance. Similarly, May et al. \cite{may2015new} designed a HDT that characterizes workers from anthropometry, functional capabilities, knowledge, skills, and expertise. However, both of them did not describe the method to measure these data in detail. Generally, the kinematic data (time, position, posture, etc.) collected by sensors are widely used in industrial production. Bilberg et al. \cite{bilberg2019digital} adopted a Kinect sensor to monitor human location and optimized robot trajectories to reduce collisions with people. Peruzzini et al. \cite{peruzzini2017benchmarking} investigated the available technologies for monitoring user experience and created a framework to monitor human health through physiological data. Shen et al. \cite{shen2017construction} trained the HDT model by using wearable sensors collecting physiological data to identify workload severity. Greco et al. \cite{greco2020digital} used the sensors to monitor the human’s kinematic motion when they were working. The model simulated by these data can be well evaluated for muscle groups fatigue of the worker, so that support assessments in work or rest schedule.

\subsection{Human Body \textbackslash Organs Modeling}\label{subsec5.2}
\subsubsection{Human organs modeling}\label{subsubsec5.2.1}
The tumor domain is the main application area of organ modeling (e.g., heart, cardiovascular and liver) of HDT. For tumor research, traditional technology can only cultivate specific tumors in vitro in the physical world, but this method cannot represent the patient's actual situation. Each modelled organ can represent the original tumour patient, which allows for simulation of an appropriate treatment plan on the HDT modelled organ. This is the main benefit of HDT organ modelling technology.

Martine et al. \cite{martinez2019cardio} proposed the Cardio Twin, a human heart HDT. Various technologies (e.g., VR, AR, and robotics) were used to demonstrate the heart model in this system. Edge computing gave subjects complete control and continuous monitoring of the information collected and processed. Mazumder et al. \cite{mazumder2019synthetic} described a cardiovascular HDT model, including blood vessels, cardiac chambers, and the central nervous system. Chakshu et al. \cite{chakshu2021towards} proposed an inverse analysis of the cardiovascular system using recurrent neural networks, to detect abdominal aortic aneurysms. Crea et al. \cite{crea2020focus} presented a cardiovascular HDT model, augmented by computer inductive (using statistical models learned from data) and deductive (incorporating multiscale knowledge a synergistic combination of mechanical modeling and simulation) to provide precision medicine for cardiovascular disease. These three papers developed HDT models to address cardiovascular issues, but with various focuses. The methods or techniques and model description are shown in Table~\ref{tab6}. Subramanian et al. \cite{subramanian2020digital} designed a liver HDT model using methods that allow nonlinear, feedback, and dynamic analyses such as ordinary differential equations. Golse et al. \cite{golse2021predicting} presented a liver HDT model to reflect the patient's status by setting the model parameters, so that the model variables were close to the measured data of each patient. The liver HDT model helped physicians in decision-making strategies Table~\ref{tab6} also analysis the difference in liver HDT between these two literatures.

\begin{table}[!ht]
\begin{center}
\begin{minipage}{\textwidth}
\caption{The analysis of human digital twin modeling for organs}\label{tab6}%
\begin{tabular}{@{}ccccl@{}}
\toprule
\multicolumn{1}{c}{Organ}  & \multicolumn{1}{c}{Refs.} & \multicolumn{1}{c}{Purpose} & \begin{tabular}[c]{@{}l@{}} Methods \textbackslash \\ Techniques \\ utilized\end{tabular} & \multicolumn{1}{c}{Model description}\\
\midrule

 

 \multirow{21}{*}{\begin{tabular}[c]{@{}c@{}}\pbox{1cm} {Cardio-\\vascular}\end{tabular}}                 
  & 
  \begin{tabular}[c]{@{}l@{}} \pbox{1cm} {Mazum- \\ der et al. \cite{mazumder2019synthetic}}\end{tabular}      
  & \begin{tabular}[c]{@{}l@{}} \pbox{3cm} {Analyzing and classifying cardiovascular disease progression.}\end{tabular}  
  & \begin{tabular}[c]{@{}l@{}} \pbox{2.5cm} {Using a bi-chamber heart, hemodynamic equations, and a baroreflex-based pressure control mechanism to generate blood pressure and flow changes.}\end{tabular}  
  & \begin{tabular}[c]{@{}l@{}} \pbox{2.6cm} {The main components of the model are the blood vessels with flow dynamics, the cardiac chambers with systolic function, and the central nervous system that regulates blood pressure.}\end{tabular} \\

  \specialrule{0em}{3pt}{3pt}
  & 
  \begin{tabular}[c]{@{}l@{}} \pbox{1cm} {Chakshu et al. \cite{chakshu2021towards}}\end{tabular}      
  & \begin{tabular}[c]{@{}l@{}} \pbox{3cm} {Making an inverse analysis of cardiovascular time series to assess the patient's cardiac status.}\end{tabular}  
  & \begin{tabular}[c]{@{}l@{}} \pbox{2.5cm} {Using Neural Networks to Help Model}\end{tabular}  
  & \begin{tabular}[c]{@{}l@{}} \pbox{2.6cm} {The one-dimensional hemodynamic model, considered systemic arteries.}\end{tabular} \\

    \specialrule{0em}{3pt}{3pt}
  & 
  \begin{tabular}[c]{@{}l@{}} \pbox{1cm} {Crea et al. \cite{crea2020focus}}\end{tabular}      
  & \begin{tabular}[c]{@{}l@{}} \pbox{3cm} {Accelerate cardiovascular research and realize the vision of precision medicine. Provide better clinical interpretability and make predictions.}\end{tabular}  
  & \begin{tabular}[c]{@{}l@{}} \pbox{2.5cm} {Using clinical data (e.g., mobile health monitor, medical images) to aid in modeling through computer-augmented inductive (statistical modeling) and deductive (mechanical modeling).}\end{tabular}  
  & \begin{tabular}[c]{@{}l@{}} \pbox{2.6cm} {Statistical and mechanical models}\end{tabular} \\

  \midrule

 
  \specialrule{0em}{2pt}{2pt}

 \multirow{7}{*}{\begin{tabular}[c]{@{}c@{}}\pbox{1cm} {Liver}\end{tabular}}                 
  & 
  \begin{tabular}[c]{@{}l@{}} \pbox{1cm} {Subra-\\manian et al. \cite{subramanian2020digital}}\end{tabular}      
  & \begin{tabular}[c]{@{}l@{}} \pbox{3cm} {Help understand liver disease and predict drug toxicity.}\end{tabular}  
  & \begin{tabular}[c]{@{}l@{}} \pbox{2.5cm} {Using a mathematical framework based on ordinary differential equations, create models represented using systems of ordinary differential equations.}\end{tabular}  
  & \begin{tabular}[c]{@{}l@{}} \pbox{2.6cm} {A system of differential equations with 112 states and hundreds of parameters is produced.}\end{tabular} \\

  & 
  \begin{tabular}[c]{@{}l@{}} \pbox{1cm} {Golse et al. \cite{golse2021predicting}}\end{tabular}      
  & \begin{tabular}[c]{@{}l@{}} \pbox{3cm} {Predicting the risk of portal hypertension after hepatectomy.}\end{tabular}  
  & \begin{tabular}[c]{@{}l@{}} \pbox{2.5cm} {Physiological data collection using devices such as sensors}\end{tabular}  
  & \begin{tabular}[c]{@{}l@{}} \pbox{2.6cm} {Physiological Model of the Liver
Model the liver as two parallel components (two half-livers)}\end{tabular} \\

\botrule
\end{tabular}
\end{minipage}
\end{center}
\end{table}

\subsubsection{Human body modeling}\label{subsubsec5.2.2}
HDT is a digital avatar of a physical human and can describe the physical human body in an all-around way in the digital world. Physical body modeling vividly presents the human body shape and appearance in HDT.

Some studies on modeling the basic information of the human body, including height, weight, and physical proportions, provided various novel schemas for body shape modeling in HDT. Firstly, parametric modeling is the most commonly used paradigm for constructing the human body shape in HDT \cite{cheng2018parametric}. For example, Cheng et al. \cite{cheng2018parametric} developed a parametric 3D modeling approach based on the background difference and the human contours separation. By improving the GoogleNet network model, they obtained both front and side views of the human body. After inputting the actual human contour, they reconstructed the human body in 3D. Secondly, the RGB-D sequences data is essential for human body modeling in HDT. Hesse et al. \cite{hesse2019learning} proposed a statistical 3D Skinned Multi-Infant Linear body model, providing a new way to build the infant body shape and movement properly. Their model had fewer input data requirements and could implement human body shape modeling based on incomplete, low-quality RGB-D sequences data, reducing the data acquisition cost. Zuo et al. \cite{zuo2020sparsefusion} developed a texture mapping method to reconstruct 3D human models based on sparse RGB-D images and successfully applied it to the personalized customization of the human body model with an error of several millimeters. Thirdly, 3D scans and 2D images can also be used to model human body shapes in HDT. Tong et al. \cite{tong2012scanning} presented a scanning system based on multiple Kinects for constructing the human body shape, which has been applied to 3D human animation and virtual try-on. Compared with the traditional 3D scanning camera, their system took less time to generate a more realistic human body for every user. Xu et al. \cite{xu2020building} proposed a critical point-based method, including KPhub-PC and KPhub-I, for estimating the high-fidelity human body model. Their method constructed the human body based on the raw human body scans and human body 2D images by KPhub-PC and KPhub-I, respectively. In addition, some researchers have mapped the human’s face onto HDT by modeling technologies. Liu et al. \cite{liu2018joint} developed a novel 3D face modeling method that combined face alignment with 3D face reconstruction, using the contour of the face in the 2D image by the regression algorithm. Lin et al. \cite{lin2022controllable} presented a reliable scheme for facial editing in video construction, further improved the generative adversarial network (GAN) so that the naturally edited face can be fused back into the video, and effectively reduced identity loss and semantic entanglement. Chu et al. \cite{chu20173d} developed a parametric face modeling framework, which gathered a large amount of data related to face parameters, then used the Kriging method to characterize the facial parameters and 3D facial synthesis. We describe the results of the above methods, which offer not only technical support for the 3D modeling of the human face, but also provide feasible solutions for various applications in HDT, such as face recognition and identity authentication.

\subsection{Human Behavior Modeling }\label{subsec5.3}

HDT behavior modeling aims to build a human behavior model in the digital world to simulate human behavior in all aspects, making HDT's behavior close to human behavior. Researchers collect human behaviors in the physical world, and endow HDT models with corresponding behaviors in the virtual world \cite{pentland1999modeling}. These behaviors include not only HDT behaviors but also interactive behaviors. According to the different modeling angles of behavioral models, we investigate behavioral models from the aspects of activity modeling, social interaction modeling, and lifestyle modeling, and study how different behaviors are modeled in HDT behavior modeling from different modeling perspectives.

\subsubsection{Activity modeling}\label{subsubsec5.3.1}
Activity modeling refers to modeling human physical activities by HDT, such as facial expression changes, running, and jumping. It is the basis of HDT behavior modeling, and the other behaviors below are extensions or combinations of these behaviors. At the same time, this is also the most basic behavioral feature of “humans” in HDT. 

Wang et al. \cite{wang2019digital} presented an approach to extract a 3D facial model from an image using a shape regression network. This method obtained high-quality 3D model pictures at minimal cost, and achieved high-accuracy facial modeling. Although this method is not strictly HDT, it is also the theoretical basis for constituting HDT. Razzaq et al. \cite{razzaq2022deepclassrooms} used Convolutional Neural Networks (CNN) to build HDT models of students’ and teachers’ facial expressions and behaviors for remote teaching and classroom supervision. This method was born out of face modeling, and on this basis, the recognition of facial expressions was added, which had higher practicability. However, the limitation of this method was that only using CNN as a modeling tool may reduce generalization, and the deployment cost was raised by the requirement of the participation of a large amount of facial expression data and the requirements on sample quality. Trobinger et al. \cite{trobinger2021dual} collected information from doctors and patients using multi-modal sensors, and established a DT of patients using motion equations. The patient’s activities and facial expressions were identified through motion analysis and support vector machines to help doctors diagnose patients remotely. This method can quickly and accurately establish a patient’s HDT model, but it cannot further learn from the patient’s historical records and medical conditions, which has certain limitations. Similarly, in the medical domain, Chen et al. \cite{chen20223d} used a 3D visualization system to build a model of knee joint motion and supported the analysis of motion trajectories to achieve real-time motion tracking. The above two methods model human motion, which is more complicated than a single part. Therefore, both these studies established a complete set of HDT models, from data collection to model presentation, which has higher usability and sets a reference standard for future HDT tasks.

\subsubsection{Social interaction modeling}\label{subsubsec5.3.2}
The social interaction model is a way to achieve interaction between HDT models, and try to simulate a classic mode of human interaction in the physical world. It requires the participation of two or more HDT in the virtual world, and they need to have interfaces that can communicate with each other \cite{miller2022unified}. It can be divided into traditional social interaction models and online social interaction models.

Traditional social interaction is the interaction between HDT and HDT through limited means, such as face-to-face communication, and telephone communication. Assadi et al. \cite{al2020user} established the lightweight Message Queuing Telemetry Transport protocol based on the “JSON” format, modeled the interaction between employers and employees in the factory, and realized a model of intelligently assigning tasks according to their conditions. This model treated social interaction as a separate module that facilitates qualitative analysis and utilization of inter-HDT interactions. Rashik Parmar et al. \cite{parmar2020building} proposed five principles for building organizational DT, and although they hardly discussed the construction of social interaction models, the corresponding rules were formulated. Organization DT was optimized by simulating human transactions, competition, etc., and realized dynamic decision-making and evolution according to the constant changes in transaction and competition behavior. This approach can serve as an excellent reference template for HDT social interaction behavior modeling.

Online social interaction modeling is another way to achieve social interaction between HDT. It is abstracted from traditional social networking. Online social networking is to communicate through online social platforms and communicate in a more accessible way. Compared with traditional social methods, online socialization is more flexible and learnable. Jianshan Sun et al. \cite{sun2021digital} observed users’ online published content and preferences. They used the pretrained Glove-vector model \cite{pennington2014glove} to transform the textinto the word vecto, built the user-likes matrix through singular value decomposition (SVD) techniques \cite{kosinski2013private} and formalized personality as an HDT model. The method of this human behavior HDT model was a multitask learning depth neural network model to predict the user’s personality through these two sources of data representation, which could improve the accuracy of personality prediction. In this study, Deep Neural Networks (DNN) outperformed the standard ML algorithm, and the best performance came from combining document and like vectors obtained with SVD.

\subsubsection{Lifestyle modeling}\label{subsubsec5.3.3}
Lifestyle modelling incorporates all of the human behaviours that take place throughout the course of a day into the HDT model by translating how people live in the physical world. This modeling method integrates the above two behaviors and has higher abstraction and universality.

Barricelli et al. \cite{barricelli2020human} established the SmartFit model, which established HDT after monitoring the athlete's behavior for a while to detect the athlete's physical condition. They treated the prediction of the athlete's behaviour as a classification problem and used the error-correcting output code (ECOC) method to address it. To impute the data and increase data reliability, they used a K-Nearest Neighbor (KNN) based method. Herrgard et al. \cite{herrgaardh2022digital} proposed a hybrid model to simulate ischemic stroke, combining formula-based modeling with ML to simulate known physiological risk factors and calculate the risk of developing the disease following intervention in real-time. This model was different from the conventional HDT model, because researchers gave the operator a model with disease risk when it was created, and asked the operator to optimize it with different interaction methods to make it unique. It was highly innovative and had high use value. In addition, R. Ferdousi et al. \cite{ferdousi2021iot} proposed an HDT of the Mental Stress model, and obtained 98\% accuracy by using ML algorithms. Data were collected from 20 sources (e.g., activity, social, lifestyle, physical and psychological data). It solved the problem that the existing stress prediction system was not suitable for dealing with various changing stress sources, and the incomplete characteristics and static prediction technology that traditional methods usually use only a single source of data (such as only wearable sensors or user devices).

\section{Human Digital Twin Application}\label{sec6}
\subsection{Healthcare Domain}\label{subsec6.1}
Healthcare is defined as a broad concept covering precision medicine \cite{ahmadi2020digital}, smart medicine, sports, health, nursing and rehabilitation \cite{jimenez2020health} etc. Healthcare refers to people's health in all aspects (e.g., physical, psychological, social levels), with the goal of "putting the individual in a state of wellbeing". One healthcare domain problem is the lack of real-time information interaction \cite{ahmadi2020digital}. Another problem is that due to the complex structure of the human body and the constant changes in the state of the human, it is difficult to extract accurate molecular data \cite{croatti2020integration}. A further issue is that different chronic diseases have a prolonged course, are frequently recurrent, progress slowly, and are expensive and difficult to detect and manage. In addition, the world is facing the problem of an aging population, so it is necessary to consider more intelligent life monitoring for the elderly \cite{ahn2020digital}\cite{gamez2020digital}\cite{vidal2020opinion}\cite{liu2019novel}.
HDT technology offers new opportunities for healthcare. It has been widely used in precision medicine and sports medicine domain \cite{erol2020digital}. It seamlessly connected the physical world and digital world, effectively integrated information \cite{el2019dtwins}, provided real-time monitoring of patient body data \cite{siemens2017digital}, enabled more real-time relevant medical responses \cite{jimenez2020health}, and improved resource utilization rate \cite{ahmadi2020digital}. Researchers also achieved predictive maintenance for the elderly \cite{vidal2020opinion} by combining HDT with healthcare, providing more accurate and faster services for elderly healthcare \cite{liu2019novel}. Furthermore, HDT is expected to achieve better disease prediction, disease diagnosis, disease treatment, continuous monitoring, medical assessment, and risk assessment without interfering with patients’ daily activities \cite{ahmadi2020digital}\cite{croatti2020integration}.

In the precision medicine domain, each person has their precision digital HDT model, which combines individual characteristics (e.g., behavior habits, lifestyle, physiological and psychological states). The customized HDT model takes an individual's unique information as an essential feature and provides a solid basis for formulating patients' personalized treatment plans. Shamanna et al. \cite{shamanna2021retrospective} designed the Twin Precision Treatment Program, a platform that used body sensors and a mobile phone application to collect data, tracked and analyzed bodily health signals and personalized treatment for each patient. Barbiero et al. \cite{barbiero2021graph} introduced a framework consisting of advanced AI methods and integrated mathematical modeling to provide a panoramic view of current and future pathophysiological conditions. In addition, HDT is expected to be widely used in tumors and surgery. Mourtzis et al. \cite{mourtzis2021smart} designed a HDT for analyzing Magnetic resonance imaging (MRI) scans, helping oncologists monitor patients’ status and predict tumor development based on biometric data. this makes possible to analyze and generate the most appropriate treatment based on a dataset of patient-related information. Ahmed et al. \cite{ahmed2021potential} discussed the potential of using HDT models in surgical learning. Although they clearly stated the assumption of model building for human self-behavior, and laid an excellent foundation for future application.

 In the sports medicine domain, HDT model not only digitizes the information related to the human body, but also constructs the human body structure with high fidelity. Various sports-related options, such as sports plans and diet, can be tested and taught on the HDT model, which dramatically reduces the risk of sports training. As a result, HDT offers a highly integrated simulation system for use in sports medicine. By creating appropriate datasets and developing appropriate simulation models, HDT model can simulate the behavior of human body and its subsystems, enabling continuous monitoring of human and prediction of their future state. Barricelli et al. \cite{barricelli2020human} established the SmartFit framework to monitor the physical condition of athletes. It built a digital model of the athletes’ bodies after inpit utting the behavior of athletes for several consecutive days, to analyze athletes' states and judge if their future exercise and diet plans are optimal. This HDT model concentrated on the behaviour of athletes and their professional conduct; social behaviour or grouping athletes for greater accuracy were not considered. However, it gave the HDT model the capacity to learn and had a high usage value. Alekseyev et al. \cite{alekseyev2021measuring} introduced a method to research mobile information and measurement systems for constructing personal dynamic portraits of the human body reflecting human movement in different domains of activity study portraits, and proposed a personalized system for assessing the recovery of the patient’s walking skills. This HDT system tracked patients’ tendencies based on assessments of temporal characteristics of walking technique, to improve motor skills during rehabilitation.

 \subsection{Industry Domain}\label{subsec6.2}
Under the trend of Industry 4.0, HDT has been given more significance. Researchers started using HDT in intelligent factories to improve productivity and production efficiency. Pires et al. \cite{pires2019digital} discussed the significance and challenges of HDT. Locklin et al. \cite{locklin2021architecture} analyzed the feasibility of HDT in the industry. They argued that HDT could adjust its relationship with the physical world at a lower cost, as natural as employee job transfers.

HDT is assigned various forms and tasks within the manufacturing industry. Because the majority of human self-behavior expansion in the manufacturing domain is the expansion of some occupational knowledge and behaviour, namely professional behaviour, researchers focus more on the expansion of human social behaviour and human self-behavior. Montini et al. \cite{montini2021meta} proposed a meta-model of factory HDT, including human psychological, physiological and behavioral aspects, which provided a reference for building a unique HDT model. Greco et al. \cite{greco2020digital} used HDT to debug human-computer interaction equipment. They took into account not only common misbehaviour but also unconscious misbehaviour when modelling misbehaviour in professional behaviour in manufacturing. By studying industry errors and using them to predict misclassification, they were able to model human error behaviour by accurately capturing actual error operations. This helped mechanical designers anticipate the potential effects of their equipment due to errors in behaviour and increased the equipment's reliability. Sun et al. \cite{sun2020digital} also adopted a similar method and improved assembly quality and efficiency by constructing HDT model to debug high-precision models. Baskaran et al. \cite{baskaran2019digital} designed HDT in automotive assembly domain to test the possible impact and labor feasibility of various solutions. Ariansyah et al. \cite{ariansyah2020towards} built HDT models by collecting human fatigue data to allocate production operations work and decision-making better. Al Assadi et al. \cite{al2020user} implemented a personnel management system for the automatic training and placement of workers by modeling employees’ behavior and professional behavior. Sharotry et al. \cite{sharotry2020digital} modeled the behavior of people carrying heavy objects and judged the risk of injury by the degree of muscle fatigue and the standard of carrying movements. They built the model by collecting the data of each joint of the person during handling and evaluating the person's physical state, and added a DT module to establish HDT model through the learning of various fatigue data to ensure the correctness of the operator's handling work. This method also focused on human self-behavior and professional behavior. However, it did not realize the steps from "general" to "specialized", because it just learned the average value of human beings through a large number of samples and made judgments with more remarkable universality. Nevertheless, it may not play a role in applying extreme cases.

 \subsection{Daily Life (General) Domain}\label{subsec6.3}

The indicators of the HDT model in the digital world should be as close as possible to human characteristics to help and guide people to work and learn better by simulating the components of the physical world. The social attributes of HDT are expected to be high. Toshima et al. \cite{toshima2020challenges} took HDT as a simulation model similar to human’s internal and external self to assist in completing daily tasks. In order to predict user weight and combine it with BMI indexes, Caballero et al. \cite{caballero2021extrapolation} created an HDT model, modelled the data obtained from the body weight sensors of people, described an algorithm for filtering and predicting human body weight, proposed various thresholds based on gender to filter possible error values. When the user was predicted to be underweight, overweight or heavier, the HDT model can provide coping suggestions on time.

HDT is also considered to be applied to the metaverse to solve problems caused by the interaction between digital HDT model and human in daily life. Stacchio et al. \cite{stacchio2022will} created the HDT model in the metaverse by scanning the human body, and asking specific questions for the HDT model to test its performance. The store clerk in the physical world accepted the request on the screen as if the HDT was visiting the store. This project studied the effect of HDT and human interaction in the context of clothing stores. In addition, preserving the identity details of real human faces plays an important role in accurately confirming the identity of the HDT model in the metaverse. Lin et al. \cite{lin2022controllable} implemented controllable face editing in video reconstruction, realized high-resolution and realistic face generation, and ensured the operability and high fidelity of HDT. This method achieved prominent identity preservation and semantic unwrapping in controllable face editing, which was superior to the most advanced method.

With the development of IoT technology, HDT is expected to achieve widely used in human smart life (e.g., smart home, smart office and smart travel) in the future. Using personal and context-related data stored and processed in HDT will bring better, more personalized, safer and more efficient applications for the intelligent environment driven by IoT devices. Zibuschka et al. \cite{zibuschka2020human} envisaged HDT application in the home, architecture and office automation domain. For example, the HDT could help observe and simulate residents' living or working behaviors in the smart home environment within a limited time.

\section{Human Digital Twin Trends and Challenges}\label{sec7}

In the above research work, technology issues related to the development of HDT are the challenges that have received the most attention in the literature. With the rapid development of advanced technology (e.g., AI, IoT, 5G \textbackslash 6G, Cloud Computing), significant progress has been made in designing HDT. Nevertheless, various challenges remain, such as data heterogeneity, data transmission and real-time interaction. After constructing an HDT, various social issues with the application of HDT in the physical world cannot be ignored. We mainly discuss three social issues, one is HDT regulation, another is ethics and privacy, the third one is trust and transparency. Finally, owing to human unique thinking and cognition, machine intelligence and cognitive simulation could be considered to optimize HDT.
 
 \subsection{Technology Issues}\label{subsec7.1}

In the aspect of data heterogeneity, HDT involves the storage and analysis of various information related to the human, such as social attributes, physical behavior, physiological signals, and so on \cite{connor2022pancreatic}. This information is not only presented in different forms (e.g., pictures, text, audio, etc.), but also stored in different systems, including Windows, Unix, Linux, and others. Multi-modal and multi-source data are used to construct an HDT. Therefore, data heterogeneity emerged, which caused difficulties in data integration and led to the “isolated information island”. Data heterogeneity hinders the further exploration and application of the inherent knowledge of the data. Although previous studies of ontology technology have not dealt with HDT, it has been widely applied in the DT to solve the problem of data heterogeneity. Singh et al. \cite{singh2021data} analyzed the challenges of data management, including data diversity, data mining, and dynamic analysis. They applied ontology technology to DT modeling to address the above challenges and proposed a DT Ontology Model. Steinmetz et al. \cite{steinmetz2018internet} used an ontology to formalize the DT in the CPS. Based on the ontology, they conducted the simulation experiment in industrial case studies and preliminarily verified that the ontology can effectively integrate multiple data sources. In short, Ontology technology is not only suitable for DT, but also hopes to be used for HDT in the future. An HDT ontology model can be constructed based on ontology technology to provide a unified paradigm for the human body information-related data, which is expected to be an effective solution for data heterogeneity \cite{radhi2022adaptive}\cite{khnaisser2022using}.

Data is crucial for the intended success of the HDT in the aspect of data transmission. The attributes of the participating physical objects, interaction data with other physical objects, and predictions of future states are typically included in data streams in the connected physical world and digital world \cite{raj2021empowering}. Blockchain technology, to a large extent, can ensure the data transmission security of HDT. And it has been widely used in DT, and has achieved excellent efficiency. A blockchain-based DT framework, named Trusted Twins for Securing Cyber-Physical Systems (TTS-CPS), was proposed by Sabah Suhail et al. \cite{suhail2022towards}, which demonstrated the feasibility and operability of the TTS-CPS for a production line in the automotive industry domain. It proved that blockchain technology successfully ensures data transmission security in DT. W. Dong et al. \cite{dong2021dual} proposed a dual-blockchain framework to improve data credibility integrated into the digital twin network (DTN), including an authorization blockchain for control operations security and a data blockchain for data content security. Therefore, blockchain technology is expected to be used in HDT to solve data transmission security. Blockchain has the priori potential to solve the security lacunae to put HDT on track.  With the decentralization and immutability characteristics of blockchain, HDT can innovate excellent \cite{raj2021empowering}.

In terms of real-time interaction, 5G technology helps HDT to address the problem of real-time interaction because the development of 5G \textbackslash 6G networks has added the additional potential for precise real-time data collection and access \cite{zhu2022survey}. Due to its high reliable connection, high speed and low delay, 5G systems have been widely applied and deployed in various countries \cite{el2022survey}. 5G is the critical technology for realizing remote real-time interaction. Various experiments verification show that the 5G video transmission delay is 49ms, which well meets the needs of real-time interaction for low delay. Although it is not widely used in the infancy stage of HDT, similarly, it can draw on the experience of the video transmission protocol based on 5G communication \cite{tang2021design} in the domain of real-time unmanned aerial vehicle video transmission. 5G has realized the gigabit speed, greater capacity and ultra-low latency, while 6G will use more advanced radio equipment and various radio waves, and can provide ultra-high speed and huge capacity in a short distance. Holographic telepresence is an essential application of 6G \cite{el2022survey}. It will allow people to communicate with the authentic 3D model of objects, bring people an immersive real-time 3D experience, and elevate human communication to a new level. With the support of holographic communication technology, the real-time and high mobility characteristic of HDT can be fully realized in the future \cite{el2022survey}.

\subsection{Social Issues}\label{subsec7.2}
At the initial stage of HDT, researchers pay more attention to the study of HDT core technology and application, however, far too little attention has been paid to the open social issues in the physical world. Thus, we notice the limitation and discuss the open issues related to the development of HDT from three terms: HDT social regulation, privacy and ethics, and trust and dependence. 

In terms of HDT social regulation, owing to HDT being widely applied in the industry, healthcare and other domain, the HDT regulatory issue requires increased attention. Robust governance mechanisms and policies are needed to oversee the ethical control and management of these digital assets \cite{kamel2021digital}. At the same time, some ethical principles in digital cloning may also apply to HDT. HDT continuously causes social regulation issues contrary to human ethics and dignity, as embodied in the United Nations Declaration on Human Cloning \cite{zhu2022survey}. It is notifiable that digital cloning challenges human ethics and social regulation. The European Patent Convention under Article 53(a) denies the grant of a patent to inventions that breach public order or morality. The European Court of Justice confirmed the need to protect human dignity and integrity, which it deems as fundamental rights \cite{truby2021human}.

In terms of privacy and ethics, first of all, data privacy and ethics issues and HDT ownership should be given priority in the construction and implementation of HDTS  \cite{lauer2022designing}. Multi-modal and multi-source data used in HDT modeling, such as patient's medical data, are compassionate and involve security and privacy issues. Theft or misuse of these data may cause serious damage to users. For traditional ML, privacy protection is supported by digital signatures and data encryption. It guarantees the protection of personal data, but causes an additional burden to the data processing and increases the processing delay \cite{chen2022decentralized}. To overcome this problem, federated learning can be considered to apply to HDT to reduce processing delay while privacy is protected. Through federated learning technology, the data does not need to be transmitted and the model training is carried out at the local client and the parameters obtained by training are uploaded to the central server finally, which reduces the risk of data leakage \cite{adnan2022federated}\cite{antunes2022federated}. For another, concerning the growing importance of data privacy and HDT ownership, ethical issues arose. The first problem is how to deal with HDT and the associated data if an (unexpected) termination of therapy arises. The majority of respondents in De Maeyer and Markopoulos \cite{de2020digital}, stated that their HDT should be deleted when they die, while others imagined it could be continuously used to promote further optimization and upgrading of the HDTS. The second problem is whether a person could be represented by his HDT. French philosopher Baudrillard proposed the fundamental ambiguity of representation is further enhanced when not a natural person but an HDT is acting on behalf. More importantly, the HDT threatens the true and false difference. It directly affects the person who is represented by an HDT \cite{braun2021represent}. Additionally, the ethical issues encountered by virtual digital humans in the metaverse are also reminiscent of HDT. The reproduced stage of a deceased artist in the form of a virtual human attracted widespread attention: Is the artist remodeling authorized? If there are interests, how should they be distributed? From a moral and ethical point of view, digital resurrection violates the deceased's dignity and may cause a series of ethical issues. It is necessary to build a metaverse with a worldview and ethical consciousness in which various HDT can live \cite{park2022metaverse}.

In terms of trust and dependence, we investigate the existing modes of human attitudes toward AI, which may be available for reference to study HDT. On the one hand, the European Commission’s High Level Expert Group adopted the position that human should establish trust relations with AI and cultivate trustworthy AI \cite{thiebes2021trustworthy}. On the other hand, Ryan et al. \cite{ryan2020ai} proposed that AI cannot be trusted, because it does not have an emotional state or can be responsible for its behavior with the emotional and normative account requirements of trust. AI meets all the requirements of rational trust, but it is not human trust, but a form of dependence. From the above two opposite views on AI trust, we summarize three points of HDT trust. (i) Multi-modal and multi-source data are used in the construction of HDT, which involves various users' privacy information. Only by protecting the ethics and privacy rights of users from infringement can make it possible to establish users’ trust in HDT. (ii) Improving computing power can significantly improve the HDT model accuracy. The HDT compute capacity requirements can be approached through High-Performance Computing (HPC) \cite{gundu2022high} and Quantum Computation \cite{massoli2022leap}. Cloud/Edge computing can further improve real-time computing by combining distributed computing with central computing \cite{hartmann2022edge}. Furthermore, the evaluation method also should be considered, and it is crucial to evaluate the prediction results to judge the classifier error and provide the reliability of the calculated recommendations \cite{barricelli2020human}. (iii) Establish a simple, operable and intelligible interactive interface. Maeyer et al. \cite{maeyer2020digital} emphasized the design of human-computer interface should be based on the important premise that users can easily understand and operate. Braun et al. \cite{braun2021represent} advised paying more attention to human-computer interfaces between human and HDT, strengthening human control of HDT and protecting people's freedom from being infringed. Fuller et al. \cite{fuller2020digital} advised explaining the HDT framework from the foundational level to help user get a deeper understanding.

\subsection{Human Thinking and Cognition Issues}\label{subsec7.3}

Human thinking and cognition are uniquely human characteristics, which greatly help people solve information-processing problems. Most of human information processing process is subconscious. Thus, it is not easy to understand people’s thoughts clearly \cite{de2014thought}. 
Aiming to make machines better serve human and improve their work efficiency, researchers devote themselves to studying machine intelligence. Originally, Turing et al. \cite{church1937turing}\cite{turing2009computing} proposed imitating human thinking through computer programs. Turing machine was first used for mathematicians processing information. It imitated human thinking, neither bionic artifacts nor imitating biological structures \cite{church1937turing}\cite{petzold2008annotated}. After Turing, various researchers began to study whether programs could ever be really intelligent, creative or even conscious \cite{simon1971human}. These studies have identified it is necessary to consider the cognitive simulation block when designing HDT system, to analyze the results of human behavior and performance in the process of processing information and improve prediction accuracy.

Cognitive simulation studies the common and individual thinking and cognitive processes of human, as well as the psychological contents, representations and constraints that determine the boundaries and forms of these processes. It analyzes how people perform intelligent tasks and uses this information to design novel technical solutions. In order to describe the interaction between people and machines, Saariluoma et al. \cite{saariluoma2020human} proposed a cognitive simulation method that used human information processes to design intelligent systems. The system analyzed human information processes, such as perception and thinking, to imitate how people process information, and designed intelligent HDT. Subsequently, they proposed a detailed HDT process model designed by using the cognitive modeling method \cite{saariluoma2021human}. They aimed to propose a practical method to design intelligent information processes from a cognitive simulation approach \cite{saariluoma2018cognitive}.

\section{Conclusions}\label{sec8}

This paper has shed more light on a cutting-edge field of study known as human digital twin (HDT). In this paper, we briefly introduce digital twin (DT) and HDT, and explore their similarities and differences. We conduct a thorough review of the literature on HDT research. There are two key points that are addressed: Along with analyzing underlying technologies, we also establish typical organizational frameworks for core HDT functions or components. In addition, we propose the HDTS architecture with the core function blocks based on the literature reviewed in this paper. Additionally, we research state-of-the-art developments in sensing, modelling the human body/ organs, and modelling human behaviour for HTD core technologies. We also present the application of HDT in the domains of healthcare, daily life and industry. Finally, we discuss various trends and challenges of future HDT development. 

A limitation of this study is that we only present the state-of-the-art in HDT technologies and applications and discuss open issues; we do not, however, propose a practical approach for generic HDT modelling technologies that would enable us to build HDT models that are appropriate for anyone regardless of location, age, gender, or other characteristics. Designing a generic model is intriguing and valuable to be usefully explored in further research. Despite its limitations, the study certainly adds to our understanding of the HDT concept, state of the art technologies, applications and challenges. In general, in the future, the rapid development of HDT will promote the further development of its core technologies, and its application will expand to more innovative domains. The open issues and challenges covered in this study show that HDT is a promising area for future research.

\bibliography{sn-article}


\end{document}